\documentstyle[12pt,epsfig]{article}
\newcommand{\agt}{\,\rlap{\lower 3.5 pt \hbox{$\mathchar \sim$}} \raise 1pt
 \hbox {$>$}\,}
\newcommand{\alt}{\,\rlap{\lower 3.5 pt \hbox{$\mathchar \sim$}} \raise 1pt
 \hbox {$<$}\,}
\newcommand{\n}{\hspace*{-2.5mm}}

\catcode`@=11
\newcount\@tempcntc
\def\@citex[#1]#2{\if@filesw\immediate\write\@auxout{\string\citation{#2}}\fi
  \@tempcnta\z@\@tempcntb\m@ne\def\@citea{}\@cite{\@for\@citeb:=#2\do
    {\@ifundefined
       {b@\@citeb}{\@citeo\@tempcntb\m@ne\@citea\def\@citea{,}{\bf ?}\@warning
       {Citation `\@citeb' on page \thepage \space undefined}}%
    {\setbox\z@\hbox{\global\@tempcntc0\csname b@\@citeb\endcsname\relax}%
     \ifnum\@tempcntc=\z@ \@citeo\@tempcntb\m@ne
       \@citea\def\@citea{,}\hbox{\csname b@\@citeb\endcsname}%
     \else
      \advance\@tempcntb\@ne
      \ifnum\@tempcntb=\@tempcntc
      \else\advance\@tempcntb\m@ne\@citeo
      \@tempcnta\@tempcntc\@tempcntb\@tempcntc\fi\fi}}\@citeo}{#1}}
\def\@citeo{\ifnum\@tempcnta>\@tempcntb\else\@citea\def\@citea{,}%
  \ifnum\@tempcnta=\@tempcntb\the\@tempcnta\else
   {\advance\@tempcnta\@ne\ifnum\@tempcnta=\@tempcntb \else \def\@citea{--}\fi
    \advance\@tempcnta\m@ne\the\@tempcnta\@citea\the\@tempcntb}\fi\fi}
\catcode`@=12

\begin{document}
\title{\vskip-3cm{\baselineskip14pt
\centerline{\normalsize DESY 98--023\hfill ISSN 0418--9833}
\centerline{\normalsize MPI/PhT/98--020\hfill}
\centerline{\normalsize hep--ph/9803256\hfill}
\centerline{\normalsize March 1998\hfill}
}
\vskip1.5cm
TEVATRON-HERA Colour-Octet Charmonium Anomaly Versus Higher-Order QCD Effects}
\author{{\sc Bernd A. Kniehl$^1$ and Gustav Kramer$^2$}\\
$^1$ Max-Planck-Institut f\"ur Physik (Werner-Heisenberg-Institut),\\
F\"ohringer Ring 6, 80805 Munich, Germany\\
$^2$ II. Institut f\"ur Theoretische Physik\thanks{Supported
by Bundesministerium f\"ur Forschung und Technologie, Bonn, Germany,
under Contract 05~7~HH~92P~(5),
and by EU Program {\it Human Capital and Mobility} through Network
{\it Physics at High Energy Colliders} under Contract
CHRX--CT93--0357 (DG12 COMA).},
Universit\"at Hamburg,\\
Luruper Chaussee 149, 22761 Hamburg, Germany}
\date{}
\maketitle
\begin{abstract}
Approximately taking into account the higher-order effects due to 
multiple-gluon initial-state radiation, we extract from the latest Tevatron
data of prompt $J/\psi$ hadroproduction the leading colour-octet matrix
elements within the nonrelativistic-QCD (NRQCD) factorization formalism
proposed by Bodwin, Braaten, and Lepage.
We find that the matrix elements which describe the formation of $J/\psi$ 
mesons from colour-octet $c\bar c$ pairs in the angular-momentum states
${}^{2S+1}\!L_J={}^1\!S_0$ and ${}^3\!P_J$, with $J=0,1,2$, which are 
responsible for the excess of the predicted cross section of inelastic
$J/\psi$ photoproduction over the existing HERA data at high values of the 
inelasticity variable $z$, are significantly reduced.
We conclude that it is premature to proclaim a discrepancy between the Tevatron
and HERA measurements of inclusive $J/\psi$ production in the NRQCD framework.
We also consider $J/\psi$ mesons originating from the radiative feed down of
promptly produced $\chi_{cJ}$ mesons.

\end{abstract}
\newpage

\section{\label{s1}Introduction}

Since its discovery in 1974, the $J/\psi$ meson has provided a useful
laboratory for quantitatively testing quantum chromodynamics (QCD) and, in
particular, the interplay of perturbative and nonperturbative phenomena.
Recently, the cross section of inclusive $J/\psi$ hadroproduction measured in
$p\bar p$ collisions at the Fermilab Tevatron \cite{abe,cdf} turned out to be
more than one order of magnitude in excess of what used to be the best
theoretical prediction \cite{bai}, based on the colour-singlet model (CSM).
As a solution to this puzzle, Bodwin, Braaten, and Lepage \cite{bod} proposed
the existence of so-called colour-octet processes to fill the gap.
The central idea is that $c\bar c$ pairs are produced at short distances in
colour-octet states and subsequently evolve into physical (colour-singlet)
charmonia by the nonperturbative emission of soft gluons.
The underlying theoretical framework is provided by nonrelativistic QCD
(NRQCD) endowed with a particular factorization hypothesis, which implies a
separation of short-distance coefficients, which are amenable to perturbative
QCD, from long-distance matrix elements, which must be extracted from
experiment.
This formalism involves a double expansion in the strong coupling constant
$\alpha_s$ and the relative velocity $v$ of the bound $c$ quarks, and takes
the complete structure of the charmonium Fock space into account.

In order to convincingly establish the phenomenological significance of the
colour-octet mechanism, it is indispensable to identify it in other kinds of
high-energy experiments as well.
In the meantime, there have been a number of interesting proposals and
encouraging results concerning possible colour-octet signatures of inclusive
$J/\psi$ production in $e^+e^-$ annihilation in the continuum \cite{che} and
on the $Z$-boson resonance \cite{keu}, hadronic collisions in the fixed-target
mode \cite{tan}, $B$-meson decays \cite{lee}, and photoproduction in the
forward direction \cite{amu} and at finite transverse momentum \cite{cac,pko}.
Although the experimental $Z$-boson-decay data from CERN LEP \cite{abr} are in
agreement with the predicted colour-octet signature, they do not exclude the
hypothesis that the observed prompt charmonium signal is produced by
colour-singlet processes alone.

In the case of inelastic $J/\psi$ photoproduction, however, NRQCD with
colour-octet matrix elements tuned \cite{cho} to fit the Tevatron data
\cite{abe} predicts \cite{cac,pko} at leading order (LO) a distinct rise in
cross section as $z\to1$, where $z$ is the fraction of the photon energy
transferred to the $J/\psi$ meson in the proton rest frame, which is not
observed by the H1 \cite{aid} and ZEUS \cite{bre} collaborations at DESY HERA.
This colour-octet charmonium anomaly has cast doubts on the validity of the
NRQCD factorization hypothesis \cite{bod}, which seems so indispensible to
interpret the Tevatron data in a meaningful way.

This paper is an attempt to rescue the NRQCD approach by approximately taking
into account dominant higher-order (HO) QCD effects.
The basic idea is as follows.
The predicted excess over the HERA data at $z$ close to unity is chiefly
generated by colour-octet $c\bar c$ pairs in the states ${}^1\!S_0$,
${}^3\!P_0$, and ${}^3\!P_2$ \cite{cac,pko}, where we use the spectroscopic
notation ${}^{2S+1}\!L_J$ to indicate the spin $S$, the orbital angular
momentum $L$, and the total angular momentum $J$.
On the other hand, in hadroproduction at the Tevatron, the contributions from
the colour-octet ${}^1\!S_0$ and ${}^3\!P_J$ states fall off much more
strongly with increasing $p_T$ than the one due to the colour-octet
${}^3\!S_1$ state \cite{cho}, which is greatly suppressed in the quasi-elastic
limit of photoproduction \cite{cac,pko}.
Consequently, the nonperturbative matrix elements which are responsible for
the colour-octet charmonium crisis are essentially fixed by the Tevatron data
in the low-$p_T$ regime.
This is precisely where the LO approximation used in Ref.~\cite{cho} is
expected to become unreliable due to multiple-gluon radiation from the
initial and final states.
In Ref.~\cite{can}, this phenomenon was carefully analyzed in a Monte Carlo
framework and found to significantly increase the LO cross section.
In this paper, we perform fits to the latest prompt $J/\psi$ data taken by the
CDF collaboration \cite{cdf} at the Tevatron incorporating this information
\cite{can} on the dominant HO QCD effects.
We then update the NRQCD predictions for inelastic $J/\psi$ photoproduction at
HERA to find out whether the colour-octet charmonium anomaly persists.
We also carry out a LO analysis for $J/\psi$ mesons originating from the
radiative feed down of promptly produced $\chi_{cJ}$ mesons.

This paper is organized as follows.
In Section~\ref{s2}, we specify the theoretical input for our analysis.
In Section~\ref{s3}, we describe our fitting procedure and present our results
for the various $J/\psi$ and $\chi_{cJ}$ matrix elements.
In Section~\ref{s4}, we present the resulting predictions for inelastic 
$J/\psi$ and $\chi_{cJ}$ photoproduction at HERA.
Our conclusions are summarized in Section~\ref{s5}.

\section{\label{s2}Theoretical input}

We now describe the theoretical framework for our analysis.
We calculate the cross section for the inclusive production $A+B\to H+X$ of
the physical charmonium state $H$ in the collision of two hadrons, $A$ and
$B$, in the parton model of QCD, i.e.\ $A$ and $B$ are represented by their
parton density functions (PDF's).
In the case of photoproduction, $A$ is a quasi-real photon, which either
directly interacts with the partons inside hadron $B$ (direct photoproduction)
or fluctuates into a bunch of quarks and gluons, which in turn interact with
the partons inside hadron $B$, while the photon remnants give rise to hadronic
activity in the original photon flight direction (resolved photoproduction).
Depending on the transverse momentum $p_T$ of the $H$ meson, we adopt two
different pictures to describe its production.
For $p_T$ values of the order of the charmonium mass $M_H\approx2m_c$, we
consider the formation of a collinear $c\bar c$ pair within the primary
hard-scattering process (fusion).
In the limit $p_T\gg M_H$, only those subprocesses survive where the $c\bar c$ 
pair is created from a single high-energy gluon, charm quark or antiquark
which is close to its mass shell (fragmentation).
It is then useful to describe this two-step process as a convolution of the
hard-scattering cross section for single-parton production with an
appropriate fragmentation function (FF).
In both cases, the factorization theorem of NRQCD \cite{bod} allows for a
systematical treatment of the transition from the $c\bar c$ pair, in state 
$n$, to the physical $H$ meson.
The states
$n=[\,\underline{1},{}^{2S+1}\!L_J],[\,\underline{8},{}^{2S+1}\!L_J]$,
where $\underline{1}$ and $\underline{8}$ indicate the colour multiplicity,
span the whole $c\bar c$ Fock space.
This formalism comprises a separation of short-distance parts, which are
amenable to perturbative QCD, from long-distance matrix elements
$\left\langle{\cal O}^H[n]\right\rangle$, which must be extracted from
experiment.
In the case of fusion, the differential cross section of $ij\to H+X$ is
decomposed as
\begin{equation}
\frac{d\sigma}{dt}(ij\to H+X)
=\sum_n\frac{d\sigma}{dt}\left(ij\to c\bar c[n]+k\right)
\left\langle{\cal O}^H[n]\right\rangle,
\label{fus}
\end{equation}
where the cross sections of the partonic subprocess $ij\to c\bar c[n]+k$ may
be calculated in NRQCD as a perturbation expansion in the strong coupling
constant $\alpha_s(\mu)$, since the renormalization scale $\mu$ is typically
set by the transverse mass $m_T=\sqrt{4m_c^2+p_T^2}$ of the $H$ meson.
In the case of fragmentation, one has
\begin{equation}
D_{i\to H}(x,\mu)
=\sum_nd_{i\to n}(x,\mu)\left\langle{\cal O}^H[n]\right\rangle,
\label{fra}
\end{equation}
where $d_{i\to n}(x,\mu)$ gives the probability for the parton $i$ to form a
jet that includes a $c\bar c$ pair in state $n$ carrying the 
longitudinal-momentum fraction $x$.
The coefficient $d_{i\to n}(x,\mu_0)$ at the initial scale $\mu_0=2m_c$
involves only momenta of order $m_c$, and can thus be calculated within NRQCD
in powers of $\alpha_s(\mu_0)$.
The evolution of the FF $D_{i\to H}(x,\mu_0)$ up to higher fragmentation
scales $\mu=M_T$ is ruled by the timelike Altarelli-Parisi (AP) equations,
which may be conveniently solved in $x$ space \cite{bin}.

The relative importance of the various terms in Eqs.~(\ref{fus}) and
(\ref{fra}) may be estimated my means of NRQCD velocity scaling rules 
\cite{bod}.
In the limit $v\to0$, where $v$ is the average velocity of the charm quark in
the $H$-meson rest frame, each of the nonperturbative matrix elements
$\left\langle{\cal O}^H[n]\right\rangle$ scales with a definite power of $v$.
Thus, Eqs.~(\ref{fus}) and (\ref{fra}) can be organized as a double expansion 
in $\alpha_s$ and $v$.
At leading order in $v$, Eqs.~(\ref{fus}) and (\ref{fra}) reduce to the 
standard factorization formulas of the colour-singlet model, which contain 
just the term referring to the state $n$ of the physical $H$ meson, i.e.\
$[\,\underline{1},{}^3\!S_1]$ and $[\,\underline{1},{}^3\!P_J]$ for the
$J/\psi$ and $\chi_{cJ}$ mesons, respectively.
In the case of the $J/\psi$ meson,
$\left\langle{\cal O}^{J/\psi}[\,\underline{1},{}^3\!S_1]\right\rangle$
is related to the nonrelativisitic radial wave function at the origin,
$R_{J/\psi}(0)$, and may thus be extracted from the measured leptonic
annihilation rate.
The QCD-improved \cite{bar} relation reads
\begin{equation}
\Gamma(J/\psi\to\ell^+\ell^-)=\frac{8\pi\alpha^2e_c^2}{9M_{J/\psi}^2}
\left\langle{\cal O}^{J/\psi}[\,\underline{1},{}^3\!S_1]\right\rangle
\left(1-\frac{16}{3}\,\frac{\alpha_s(M_{J/\psi})}{\pi}\right),
\label{gee}
\end{equation}
where $\alpha$ is Sommerfeld's fine-structure constant and $e_c=2/3$ is the
fractional charge of the charm quark.
In a similar way,
$\left\langle{\cal O}^{\chi_{cJ}}[\,\underline{1},{}^3\!P_J]\right\rangle$ may
be determined from the measured partial widths of the $\chi_{cJ}$-meson decays
into light hadrons \cite{gtb} or two photons \cite{hua}.
The leading colour-octet matrix elements of the $J/\psi$ meson are
$\left\langle{\cal O}^{J/\psi}[\,\underline{8},{}^1\!S_0]\right\rangle$,
$\left\langle{\cal O}^{J/\psi}[\,\underline{8},{}^3\!S_1]\right\rangle$, and
$\left\langle{\cal O}^{J/\psi}[\,\underline{8},{}^3\!P_J]\right\rangle$, with
$J=0,1,2$.
The leading colour-octet matrix element of the $\chi_{cJ}$ meson is
$\left\langle{\cal O}^{\chi_{cJ}}[\,\underline{8},{}^3\!S_1]\right\rangle$.
Due to heavy-quark spin symmetry, the $J$-dependent matrix elements satisfy
the multiplicity relations
\begin{eqnarray}
\left\langle{\cal O}^{J/\psi}[\,\underline{8},{}^3\!P_J]\right\rangle
&\n=\n&(2J+1)
\left\langle{\cal O}^{J/\psi}[\,\underline{8},{}^3\!P_0]\right\rangle,
\nonumber\\
\left\langle{\cal O}^{\chi_{cJ}}[\,\underline{1},{}^3\!P_J]\right\rangle
&\n=\n&(2J+1)
\left\langle{\cal O}^{\chi_{c0}}[\,\underline{1},{}^3\!P_0]\right\rangle,
\nonumber\\
\left\langle{\cal O}^{\chi_{cJ}}[\,\underline{8},{}^3\!S_1]\right\rangle
&\n=\n&(2J+1)
\left\langle{\cal O}^{\chi_{c0}}[\,\underline{8},{}^3\!S_1]\right\rangle,
\end{eqnarray}
up to terms of relative order $v^2$.
The scaling with $v$ (and $m_c$) of the various $J/\psi$ and $\chi_{cJ}$
matrix elements is indicated in Tables~\ref{t1} and \ref{t2}, respectively.

The leading colour-octet matrix elements for the $J/\psi$ and $\chi_{cJ}$ 
mesons have been determined through fits, based on the fusion \cite{cho,ben}
and fragmentation \cite{fle} mechanisms at LO, to CDF data on inclusive
charmonium hadroproduction \cite{abe,cdf}.
In this paper, we perform new fits to the most recent CDF data \cite{cdf}
taking into account information on the HO QCD effects induced by 
multiple-gluon radiation in a Monte Carlo framework \cite{can} and
consistently combining the fusion and fragmentation pictures.

The hadroproduction or resolved photoproduction of charmonium at finite $p_T$
via fusion proceeds at LO via the subprocesses $gg\to c\bar c+g$,
$gq\to c\bar c+q$, and $q\bar q\to c\bar c+g$, where $q$ stands for a light
quark or antiquark.
The cross sections for the colour-singlet states $[\,\underline{1},{}^3\!S_1]$
and $[\,\underline{1},{}^3\!P_J]$ may be found in Ref.~\cite{bai} and those
for the colour-octet states $[\,\underline{8},{}^1\!S_0]$,
$[\,\underline{8},{}^3\!S_1]$, and $[\,\underline{8},{}^3\!P_J]$ in
Ref.~\cite{cho}.
As for direct photoproduction, the contributing subprocesses are
$\gamma g\to c\bar c+g$ and $\gamma q\to c\bar c+q$, where the photon can
either couple to the $q$- or $c$-quark line.
The cross sections for the colour-singlet and most of the colour-octet states
are presented in Refs.~\cite{ber} and \cite{pko}, respectively.
The cross section of $\gamma g\to c\bar c[\,\underline{8},{}^3\!S_1]+g$, which
is not explicitly listed in Ref.~\cite{pko}, is obtained from the one of
$\gamma g\to c\bar c[\,\underline{1},{}^3\!S_1]+g$ through multiplication with
15/8.
At LO, $[\,\underline{1},{}^3\!S_1]$ can only be produced through
$gg\to c\bar c+g$ and $\gamma g\to c\bar c+g$.
Furthermore, $[\,\underline{1},{}^3\!P_J]$ cannot be generated via
$\gamma g\to c\bar c+g$ and $\gamma q\to c\bar c+q$, i.e.\ the direct 
photoproduction of $\chi_{cJ}$ mesons is a pure colour-octet process.

The direct and resolved photoproduction of $J/\psi$ and $\chi_{cJ}$ mesons via
fragmentation was extensively discussed in Refs.~\cite{god,kni,bak}.
The analysis of resolved photoproduction \cite{kni,bak} may be readily
converted to hadroproduction by replacing the photon PDF's with the respective
proton PDF's.
The relevant partonic cross sections are available at next-to-leading order
(NLO) in massless QCD \cite{aur}.
The $d_{i\to n}$ functions in Eq.~(\ref{fra}) pertinent to $i=g,c,\bar c$ and
$n=[\,\underline{1},{}^3\!S_1],[\,\underline{1},{}^3\!P_J],
[\,\underline{8},{}^3\!S_1]$ were found in Ref.~\cite{bra} and collected in
the Appendix of Ref.~\cite{kni}.
Since the gluon is a $[\,\underline{8},{}^3\!S_1]$ state,
$d_{g\to [\,\underline{8},{}^3\!S_1]}$ is of ${\cal O}(\alpha_s)$, while
$d_{g\to [\,\underline{8},{}^1\!S_0]}$ and
$d_{g\to [\,\underline{8},{}^3\!P_J]}$ require the emission of at least one
extra gluon and are thus of ${\cal O}(\alpha_s^2)$ and beyond.
As $\left\langle{\cal O}^{J/\psi}[\,\underline{8},{}^1\!S_0]\right\rangle$,
$\left\langle{\cal O}^{J/\psi}[\,\underline{8},{}^3\!S_1]\right\rangle$, and
$\left\langle{\cal O}^{J/\psi}[\,\underline{8},{}^3\!P_J]\right\rangle$ all
scale as $v^7$, it hence follows that the fragmentation process
$g\to c\bar c[\,\underline{8},{}^3\!S_1]$ dominates, while the other
gluon-initiated colour-octet contributions are marginal.
The $d_{i\to n}$ functions for $i=c,\bar c$ are of ${\cal O}(\alpha_s^2)$, so 
that the colour-octet contributions due to charm fragmentation are suppressed
by $v^4$ relative to the corresponding colour-singlet contributions.
In contrast to the fusion picture, it is therefore justified to disregard the
$[\,\underline{8},{}^1\!S_0]$ and $[\,\underline{8},{}^3\!P_J]$ states in the
fragmentation picture, as was done in Refs.~\cite{god,kni,bak}.

We take the renormalization scale $\mu$ and the common factorization scale
$M_f$ to be $\mu=M_f=m_T$, where $m_T$ is defined above Eq.~(\ref{fra}).
We define the starting scale $\mu_0$ of the FF's as $\mu_0=2m_c=M_{J/\psi}$.
For our LO analysis, we choose CTEQ4L \cite{lai} and GRV-LO \cite{glu} as the
proton and photon PDF's, respectively, and evaluate $\alpha_s$ from the
one-loop formula with $\Lambda^{(4)}=236$~MeV \cite{lai}.
Whenever we work at NLO, we adopt the $\overline{\rm MS}$ renormalization and
factorization scheme and employ CTEQ4M \cite{lai}, GRV-HO \cite{glu}, and the
two-loop formula for $\alpha_s$ with
$\Lambda_{\overline{\rm MS}}^{(4)}=296$~MeV \cite{lai}.
In the case of fragmentation, we start from the NLO partonic cross sections
calculated in the $\overline{\rm MS}$ scheme with $n_f=4$ massless flavours
\cite{aur}, and modify the factorization scheme for the collinear
singularities associated with final-state charm quarks so as to match the
finite-$m_c$ calculation.
This procedure, which was proposed in Ref.~\cite{spi}, is equivalent to the 
matching between the massless-charm calculation in connection with the
perturbative FF's of Ref.~\cite{mel} and the massive-charm calculation without
FF's.

Unfortunately, not all ingredients which would be necessary for a fully
consistent NLO analysis are yet available.
In the case of fusion, the NLO corrections to the cross sections
$(d\sigma/dt)\left(ij\to c\bar c+k\right)$ in Eq.~(\ref{fus}) are only known
for direct photoproduction in the CSM \cite{kra}.
Furthermore, in the case of fragmentation, the NLO corrections to the
$d_{i\to n}$ functions in Eq.~(\ref{fra}) are still unknown.

In Ref.~\cite{kra}, it was found that, under typical HERA conditions, the bulk
of the NLO corrections to the inclusive cross section of direct $J/\psi$
photoproduction in the CSM is due to the shift in
$\left\langle{\cal O}^{J/\psi}[\,\underline{1},{}^3\!S_1]\right\rangle$ which
occurs if the QCD correction ($K$) factor \cite{bar} is included in
Eq.~(\ref{gee}), while the genuine NLO corrections arising from the partonic
cross sections in Eq.~(\ref{fus}) and the proton PDF's only amount to about
20\% in the inelastic regime $z\alt0.9$.
Here, $z$ is the inelasticity variable defined as
$z=p_p\cdot p_{J/\psi}/p_p\cdot p_\gamma$, where $p_p$, $p_\gamma$, and
$p_{J/\psi}$ are the proton, photon, and $J/\psi$ four-momenta, respectively.
In our recent study of inclusive $J/\psi$ photoproduction via fragmentation
\cite{kni}, where we included both the direct and resolved channels, we found
$K$ factors of about 0.9 and 1.8 for the colour-singlet and colour-octet
processes, respectively.
Guided by the observations made in Ref.~\cite{kra}, we do not expect the
genuine NLO corrections to the $d_{i\to n}$ functions in Eq.~(\ref{fra}),
which are still missing, to drastically change these findings.
Vice versa, we infer from the fragmentation results of Ref.~\cite{kni} that,
in the case of fusion, the unknown $K$ factors for direct photoproduction via
the colour-octet mechanism and for resolved photoproduction via the
colour-singlet and colour-octet mechanisms should also be of perturbative 
magnitude.
In the case of direct photoproduction via the colour-octet mechanism, this
expectation is also supported by a recent NLO calculation of the total cross 
section \cite{pet}.
In fact, at the typical photon-proton CM energy $W=100$~GeV, the $K$ factors
in the $[\,\underline{8},{}^1\!S_0]$, $[\,\underline{8},{}^3\!P_0]$, and
$[\,\underline{8},{}^3\!P_2]$ channels turned out to be as low as 1.26, 1.05,
and 1.31, respectively \cite{pet}.
We conclude that, for inclusive charmonium photoproduction at HERA, the
present uncertainty in the size of the colour-octet matrix elements is likely
to be more significant than the partial lack of knowledge of the genuine NLO
corrections to the partonic cross sections in Eq.~(\ref{fus}) and the
$d_{i\to n}$ functions in Eq.~(\ref{fra}).

The situation should be very different for inclusive charmonium
hadroproduction at the Tevatron, especially in the low-$p_T$ range, where one
expects substantial HO QCD effects due to multiple-gluon radiation.
Such effects were estimated for the fusion mechanism in Ref.~\cite{can} by
means of the Monte Carlo event generator PYTHIA \cite{sjo} after implementing
therein the relevant colour-octet processes, and they were indeed found to be
very sizeable.
The resulting $K$ factor for the $p_T$ distribution of prompt $J/\psi$ mesons
may be extracted from Fig.~1 of Ref.~\cite{can} and is conveniently
parameterized as
\begin{equation}
K(p_T)=2.88-2.20\cdot\left(\frac{p_T}{10~{\rm GeV}}-1\right)
+1.04\cdot\left(\frac{p_T}{10~{\rm GeV}}-1\right)^2.
\label{kfa}
\end{equation}
It ranges between 4 and 2 for 5~GeV${}<p_T<{}$15~GeV.
We assume that this $K$ factor also approximately applies to the
$[\,\underline{1},{}^3\!S_1]$, $[\,\underline{8},{}^1\!S_0]$,
$[\,\underline{8},{}^3\!S_1]$, and $[\,\underline{8},{}^3\!P_J]$ channels
separately.
Thus, in order to estimate the HO-improved cross section of inclusive $J/\psi$ 
hadroproduction via fusion, we consistently evaluate these channels at LO and
in turn multiply them by the $K$ factor of Eq.~(\ref{kfa}).
On the other hand, our approximate NLO treatment of hadroproduction via
fragmentation emerges from the corresponding analysis of resolved
photoproduction \cite{kni,bak} by replacing the photon PDF's with proton
PDF's.
If we extrapolate Eq.~(\ref{kfa}) to $p_T=20$~GeV, we obtain $K=1.72$.
This result is comparable to the value 1.45, which we find for the dominant
fragmentation channel, $[\,\underline{8},{}^3\!S_1]$, using the same value
for $\left\langle{\cal O}^{J/\psi}[\,\underline{8},{}^3\!S_1]\right\rangle$
both at LO and NLO.
This indicates that the duality of the fusion and fragmentation pictures also
carries over to higher orders.

\section{\label{s3}Fit to the data of charmonium hadroproduction}

We now describe our fitting procedure.
As experimental input, we use the latest CDF data samples of prompt $J/\psi$
mesons and of $J/\psi$ mesons originating from the radiative decays
$\chi_{cJ}\to J/\psi+\gamma$ of prompt $\chi_{cJ}$ mesons \cite{cdf}.
The $J/\psi$ mesons were detected via their $J/\psi\to\mu^+\mu^-$ decays.
The data were collected in $p\bar p$ collisions with centre-of-mass (CM)
energy $\sqrt s=1.8$~TeV at the Fermilab Tevatron and come as the differential
cross section $d\sigma/dp_T$ integrated over the rapidity range $|\eta|<0.6$.
Each data sample consists of 11 data points ranging from $p_T=5.24$~GeV to
18.38~GeV.
We adopt the measured values of the branching fractions
$B(\chi_{cJ}\to J/\psi+\gamma)$ and $B(J/\psi\to\mu^+\mu^-)$ from 
Ref.~\cite{pdg}.
We determine
$\left\langle{\cal O}^{J/\psi}[\,\underline{1},{}^3\!S_1]\right\rangle$ from
the experimental value of $\Gamma(J/\psi\to\ell^+\ell^-)$ \cite{pdg} via
Eq.~(\ref{gee}).
Our LO and NLO results are listed in Table~\ref{t1}.

We first concentrate on prompt $J/\psi$ hadroproduction.
In Refs.~\cite{cho,ben}, the $J/\psi$ colour-octet matrix elements were fitted
on the basis of the fusion picture at LO, whereby the
$[\,\underline{8},{}^3\!S_1]$ component was supplemented with leading
logarithms from $g\to c\bar c[\,\underline{8},{}^3\!S_1]$ fragmentation.
This was achieved in an ad-hoc way, namely by multiplication with the ratio
of the corresponding gluon-fragmentation cross sections with and without LO AP
evolution.
Here, we adopt a more rigorous approach.
We introduce a variable separation cut $p_T^{\rm cut}$, which we choose to
coincide with any of the 11 CDF data points, as a demarcation between the
fusion and fragmentation pictures.
In the first step, we determine
$\left\langle{\cal O}^{J/\psi}[\,\underline{8},{}^3\!S_1]\right\rangle$ by
fitting the data points for $p_T\ge p_T^{\rm cut}$ in the fragmentation 
picture.
This result is then used as input for the second step, in which
$\left\langle{\cal O}^{J/\psi}[\,\underline{8},{}^1\!S_0]\right\rangle$ and
$\left\langle{\cal O}^{J/\psi}[\,\underline{8},{}^3\!P_0]\right\rangle$ are
determined by fitting the data points for $p_T\le p_T^{\rm cut}$ in the
fusion picture.
As was noticed in Ref.~\cite{cho}, the cross sections of
$c\bar c[\,\underline{8},{}^1\!S_0]$ and $c\bar c[\,\underline{8},{}^3\!P_0]$
production have very similar $p_T$ dependences, so that treating
$\left\langle{\cal O}^{J/\psi}[\,\underline{8},{}^1\!S_0]\right\rangle$ and
$\left\langle{\cal O}^{J/\psi}[\,\underline{8},{}^3\!P_0]\right\rangle$ as 
independent fit parameters would yield highly correlated results.
It was therefore suggested \cite{cho} to fix a specific linear combination of
$\left\langle{\cal O}^{J/\psi}[\,\underline{8},{}^1\!S_0]\right\rangle$ and
$\left\langle{\cal O}^{J/\psi}[\,\underline{8},{}^3\!P_0]\right\rangle$ 
instead.
Following this suggestion, we define 
\begin{equation}
M_r^{J/\psi}=
\left\langle{\cal O}^{J/\psi}[\,\underline{8},{}^1\!S_0]\right\rangle
+\frac{r}{m_c^2}
\left\langle{\cal O}^{J/\psi}[\,\underline{8},{}^3\!P_0]\right\rangle,
\label{mrj}
\end{equation}
where $r$ is to be chosen in such a way that the superposition of these two 
channels is insensitive to precisely how they are weighted relative to each 
other.
We determine $r$ to be the ratio of the fit result for
$\left\langle{\cal O}^{J/\psi}[\,\underline{8},{}^1\!S_0]\right\rangle$ under 
the condition
$\left\langle{\cal O}^{J/\psi}[\,\underline{8},{}^3\!P_0]\right\rangle=0$
to the one for
$\left\langle{\cal O}^{J/\psi}[\,\underline{8},{}^3\!P_0]\right\rangle/m_c^2$
under the condition
$\left\langle{\cal O}^{J/\psi}[\,\underline{8},{}^1\!S_0]\right\rangle=0$.
We fit by analytically minimizing the $\chi^2$ values.
We measure the overall quality of the fit by adding the two $\chi^2$ values
achieved in the determination of
$\left\langle{\cal O}^{J/\psi}[\,\underline{8},{}^3\!S_1]\right\rangle$
in the upper $p_T$ range and of $M_r^{J/\psi}$ in the lower $p_T$ range.
Finally, we determine $p_T^{\rm cut}$ by minimizing the total $\chi^2$.
It turns out that the best results are obtained if the first 10 (last 2) data
points are described according to the fusion (fragmentation) picture, i.e.\
the transition between these two pictures happens somewhere around
$p_T=15$~GeV.

\begin{table}
\begin{center}
\caption{Values of the $J/\psi$ matrix elements and of $r$ resulting from the
LO and HO-improved fits to the CDF data \protect\cite{cdf}.
$M_r^{J/\psi}$ is defined in Eq.~(\ref{mrj}).
The first 10 (last 2) data points are described in the fusion (fragmentation)
picture.}
\label{t1}
\smallskip
\begin{tabular}{|c|c|c|c|}
\hline\hline
 & LO & HO & scaling \\
\hline
$\left\langle{\cal O}^{J/\psi}[\,\underline{1},{}^3\!S_1]\right\rangle$ &
$(7.63\pm0.54)\cdot10^{-1}$~GeV$^3$ & $(1.30\pm0.09)$~GeV$^3$ &
$[m_c^3v^3]$ \\
$\left\langle{\cal O}^{J/\psi}[\,\underline{8},{}^3\!S_1]\right\rangle$ &
$(3.94\pm0.63)\cdot10^{-3}$~GeV$^3$ & $(2.73\pm0.45)\cdot10^{-3}$~GeV$^3$ &
$[m_c^3v^7]$ \\
$M_r^{J/\psi}$ &
$(6.52\pm0.67)\cdot10^{-2}$~GeV$^3$ & $(5.72\pm1.84)\cdot10^{-3}$~GeV$^3$ &
$[m_c^3v^7]$ \\
$r$ & 3.47 & 3.54 & \\
$\chi_{\rm DF}^2$ fus.\ & 5.97/10 & 2.23/10 & \\
$\chi_{\rm DF}^2$ fra.\ & 1.53/2 & 1.73/2 & \\
$\chi_{\rm DF}^2$ tot.\ & 7.49/12 & 3.96/12 & \\
\hline\hline
\end{tabular}
\end{center}
\end{table}

Figures~\ref{f1}(a) and (b) illustrate for the LO and HO-improved analyses,
respectively, how the fusion and fragmentation cross section compare with the
experimental data and how they are decomposed into their colour-singlet and
colour-octet components.
The fragmentation results are only shown for $p_T\ge10.91$~GeV, while the 
fusion results are displayed over the full $p_T$ range.
In each case, the total result (solid lines) is obtained as the superposition
of the colour-singlet (dotted lines) and colour-octet contributions.
The $[\,\underline{8},{}^3\!S_1]$ contributions (dashed lines) of the fusion 
and fragmentation pictures almost coincide.
The colour-octet contribution proportional to $M_r^{J/\psi}$ (dot-dashed line)
is only included in the fusion picture, while it is neglected in the
fragmentation picture for reasons explained in Section~\ref{s2}.
Notice that the LO $[\,\underline{1},{}^3\!S_1]$ contributions in the
fusion and fragmentation pictures arise from very different sources.
On the one hand, the LO fusion subprocess $gg\to c\bar c+g$ has no counterpart
in the fragmentation picture.
On the other hand, the LO fragmentation processes $g\to c\bar c+gg$,
$c\to c\bar c+c$, and $\bar c\to c\bar c+\bar c$ correspond to HO
contributions in the fusion picture.
This explains why the dotted lines in Figs.~\ref{f1}(a) and (b) do not match.
As is by now common knowledge, the colour-octet processes are indeed necessary
in order to reconcile theory with experiment as far as inclusive
hadroproduction of prompt charmonium at the Tevatron is concerned.
The results and the various $\chi^2$ values of the LO and HO-improved fits are
summarized in Table~\ref{t1}.
The errors quoted in Table~\ref{t1} only include the experimental errors on
the CDF data.
In addition, there are the usual theoretical uncertainties due to the 
dependence on the renormalization and factorization scales and on the proton
PDF's, which have been carefully estimated in Ref.~\cite{ben}.
Our LO results should be compared with those obtained in Refs.~\cite{cho,ben}.
We find reasonable agreement for $M_r^{J/\psi}$ (and $r$), which was
determined to be $M_3^{J/\psi}=(6.6\pm1.5)\cdot10^{-2}$~GeV$^3$ in 
Ref.~\cite{cho} and $M_{3.5}^{J/\psi}=(4.38\pm1.15)\cdot10^{-2}$~GeV$^3$ (for
CTEQ4L \cite{lai}) in Ref.~\cite{ben}.
On the other hand, our result for
$\left\langle{\cal O}^{J/\psi}[\,\underline{8},{}^3\!S_1]\right\rangle$ is
somewhat smaller than the value $(6.6\pm2.1)\cdot10^{-3}$~GeV$^3$ of
Ref.~\cite{cho} and considerably smaller than the value
$(1.06\pm0.14)\cdot10^{-2}$~GeV$^3$ of Ref.~\cite{ben}.
We attribute this deviation to the different implementations of the
fragmentation contribution.
As we pass from LO to our approximate HO implementation, we observe that
$\left\langle{\cal O}^{J/\psi}[\,\underline{8},{}^3\!S_1]\right\rangle$ is
only slightly decreased, by 31\%, while $M_r^{J/\psi}$ drops off by more than 
one order of magnitude.
The change in
$\left\langle{\cal O}^{J/\psi}[\,\underline{8},{}^3\!S_1]\right\rangle$
precisely compensates the enhancement due to the $K$ factor in the
$g\to c\bar c[\,\underline{8},{}^3\!S_1]$ fragmentation channel at high $p_T$.
The dramatic reduction of $M_r^{J/\psi}$ is due to the interplay of two 
effects.
On the one hand, the $[\,\underline{8},{}^3\!S_1]$ fusion channel, whose
matrix element is fixed by the fit in the fragmentation regime, is strongly
enhanced at low $p_T$ due to multiple-gluon radiation according to
Eq.~(\ref{kfa}) and only leaves little room for possible contributions due
the $[\,\underline{8},{}^1\!S_0]$ and $[\,\underline{8},{}^3\!P_J]$ fusion 
channels.
On the other hand, the latter channels are also strongly enhanced at low $p_T$
by the $K$ factor of Eq.~(\ref{kfa}), so that $M_r^{J/\psi}$ is even further
reduced.
The situation is nicely illustrated if we compare the relative importance of 
the various colour-octet fusion contributions in Figs.~\ref{f1}(a) and (b).
At LO, the combined $[\,\underline{8},{}^1\!S_0]$ and
$[\,\underline{8},{}^3\!P_J]$ contributions exceed the one due to the
$[\,\underline{8},{}^3\!S_1]$ channel way up to $p_T=10$~GeV, while, at HO,
the $[\,\underline{8},{}^3\!S_1]$ contribution is dominant over the full $p_T$ 
range considered.
Our HO-improved results for
$\left\langle{\cal O}^{J/\psi}[\,\underline{8},{}^3\!S_1]\right\rangle$ and
$M_r^{J/\psi}$ are very similar to those extracted in Ref.~\cite{can} from the
previous CDF data \cite{abe} with the 1994 GRV-HO proton PDF's \cite{grv},
which correctly account for the low-$x$ behaviour of the proton structure
function $F_2(x,Q^2)$ measured at HERA, namely
$\left\langle{\cal O}^{J/\psi}[\,\underline{8},{}^3\!S_1]\right\rangle
=(3.4\pm0.4)\cdot10^{-3}$~GeV$^3$ and
$M_3^{J/\psi}=(6.0\pm1.2)\cdot10^{-3}$~GeV$^3$.
Finally, we note that the overall $\chi^2$ per degree of freedom
($\chi_{\rm DF}^2$) is significantly reduced, from 0.62 to 0.33, as we pass
from LO to HO.

\begin{table}
\begin{center}
\caption{Values of the $\chi_{cJ}$ matrix elements resulting from the LO fit
to the CDF data \protect\cite{cdf}.
All 11 data points are described in the fusion picture.}
\label{t2}
\smallskip
\begin{tabular}{|c|c|c|c|}
\hline\hline
 & LO & scaling \\
\hline
$\left\langle{\cal O}^{\chi_{c0}}[\,\underline{1},{}^3\!P_0]\right\rangle$ &
$(2.29\pm0.25)\cdot10^{-1}$~GeV$^5$ & $[m_c^5v^5]$ \\
$\left\langle{\cal O}^{\chi_{c0}}[\,\underline{8},{}^3\!S_1]\right\rangle$ &
$(6.81\pm1.75)\cdot10^{-4}$~GeV$^3$ & $[m_c^3v^5]$ \\
$\chi_{\rm DF}^2$ fus.\ & 3.82/11 & \\
\hline\hline
\end{tabular}
\end{center}
\end{table}

In the remainder of this section, we discuss the extraction of the $\chi_{cJ}$
matrix elements through a fit to the CDF data on prompt $\chi_{cJ}$ mesons
\cite{cdf}.
For simplicity, we work at LO in the fusion picture.
The fit results for
$\left\langle{\cal O}^{\chi_{c0}}[\,\underline{1},{}^3\!P_0]\right\rangle$ and
$\left\langle{\cal O}^{\chi_{c0}}[\,\underline{8},{}^3\!S_1]\right\rangle$ are
displayed in Table~\ref{t2}.
The quality of the fit is rather high, with $\chi_{\rm DF}^2=0.35$.
Our value for
$\left\langle{\cal O}^{\chi_{c0}}[\,\underline{1},{}^3\!P_0]\right\rangle$ is 
somewhat larger than those extracted from the partial widths of the
$\chi_{cJ}$-meson decays to light hadrons \cite{gtb} and two photons
\cite{hua}.
In turn, our value for
$\left\langle{\cal O}^{\chi_{c0}}[\,\underline{8},{}^3\!S_1]\right\rangle$ is
smaller than the one found in Ref.~\cite{cho}.
The interplay of the colour-singlet and color-octet processes and the 
comparison of their superposition with the CDF data is illustrated in
Fig.~\ref{f2}.

\section{\label{s4}Predictions for charmonium photoproduction}

Having extracted the leading $J/\psi$ and $\chi_{cJ}$ colour-octet matrix
elements from the latest CDF data taking into account information on HO QCD
effects, we are now in a position to explore the phenomenological consequences
for inclusive charmonium photoproduction in $ep$ collisions at HERA.
HERA is presently operated in such a way that $E_e=27.5$~GeV positrons collide
with $E_p=820$~GeV protons in the laboratory frame, so that approximately
$\sqrt s=300$~GeV is available in the CM system.
In the case of photoproduction, the positrons act as a source of energetic,
quasi-real photons, whose energy distribution is well described in the
Weizs\"acker-Williams approximation \cite{wei} by a well-known formula, which
may be found e.g.\ in Eq.~(17) of Ref.~\cite{kni}.

The H1 \cite{aid} and ZEUS \cite{bre} collaborations recently presented their
measurements of inelastic $J/\psi$ photoproduction.
In both experiments, the scattered positron was not tagged, so that the
maximum photon virtuality was $Q_{\rm max}^2=4$~GeV$^2$.
The measured $ep$ cross sections were converted to averaged $\gamma p$ cross
sections by dividing out the photon-flux factor, which was evaluated by
integrating the Weizs\"acker-Williams formula over the considered 
photon-energy range.
The experimental information was presented in the form of distributions
differential in the photon-proton CM energy $W$, the $J/\psi$ transverse
momentum $p_T$, and the inelasticity variable $z$.
We adhere to the kinematic cuts used in the ZEUS analysis, namely
$0.4<z<0.8$ and $p_T^2>1$~GeV$^2$ for the $W$ distribution,
$0.4<z<0.9$ and 50~GeV${}<W<180$~GeV for the $p_T^2$ distribution, and
$p_T^2>1$~GeV$^2$ and 50~GeV${}<W<180$~GeV for the $z$ distribution.
The photon-flux factor corresponding to this $W$ interval is 0.106.
By contrast, the H1 data refer the $W$ interval 30~GeV${}<W<150$~GeV, for
which the flux factor is 0.150.
Actually, the ZEUS measurement only reached down to $z>0.4$ and was
extrapolated to $z=0$ by Monte Carlo simulations yielding an enhancement
factor of 1.10 \cite{bre}.
The H1 data were treated in a similar way.
We undo this artificial extrapolation and divide the published $W$ and $p_T^2$
distributions \cite{aid,bre} by this factor.
The contribution due to $\psi^\prime$ mesons with subsequent decay into
$J/\psi$ mesons was not subtracted from the ZEUS data.
As in Ref.~\cite{bre}, we thus multiply the theoretical predictions by an 
overall factor of 1.15 to approximately account for this contribution.
The data are mostly concentrated in the low-$p_T$ range, where the fusion
picture should be valid.

In Figs.~\ref{f3}(a)--(c), we compare our LO predictions for the $W$, $p_T^2$,
and $z$ distributions, respectively, with the ZEUS and H1 data.
The circumstance that 
$\left\langle{\cal O}^{J/\psi}[\,\underline{8},{}^1\!S_0]\right\rangle$ and
$\left\langle{\cal O}^{J/\psi}[\,\underline{8},{}^3\!P_0]\right\rangle$
are not separately fixed by the fit to the CDF data induces some uncertainty 
in the colour-octet contributions to the cross sections of direct and resolved
photoproduction and thus also in the total cross section.
This uncertainty is encompassed by the results for
$\left\langle{\cal O}^{J/\psi}[\,\underline{8},{}^1\!S_0]\right\rangle=
M_r^{J/\psi}$ and
$\left\langle{\cal O}^{J/\psi}[\,\underline{8},{}^3\!P_0]\right\rangle=0$ and
those for
$\left\langle{\cal O}^{J/\psi}[\,\underline{8},{}^1\!S_0]\right\rangle=0$ and
$\left\langle{\cal O}^{J/\psi}[\,\underline{8},{}^3\!P_0]\right\rangle
=(m_c^2/r)M_r^{J/\psi}$, which are actually shown in Figs.~\ref{f3}(a)--(c).
From Fig.~\ref{f3}(c), we observe that the colour-octet contribution of
direct photoproduction is dominant for $z\agt0.5$.
Thus, it also makes up the bulk of the cross sections shown in
Fig.~\ref{f3}(a) and (b), which are integrated over $0.4<z<0.8$ and
$0.4<z<0.9$, respectively.
As is evident from Figs.~\ref{f3}(b) and (c), this contribution is also
responsible for the significant excess of the LO predictions over the
experimental results for $d\sigma/dp_T^2$ and $d\sigma/dz$ at low $p_T$ and
high $z$, respectively.
(By contrast, such an excess does not show up in Fig.~\ref{f3}(a), where the
$z$ integration only extends up to 0.8.)
This observation was assessed in the ZEUS publication \cite{bre} by the
statement that {\it the predictions of a specific leading-order colour-octet
model, as formulated to describe the CDF data on $J/\psi$ hadroproduction, are
not consistent with the data.}
To our mind, the salient point is that those predictions were made on the 
basis of LO fits to the CDF data.
It is crucial to notice that the colour-octet contribution of direct
photoproduction is essentially due the $[\,\underline{8},{}^1\!S_0]$,
$[\,\underline{8},{}^3\!P_0]$, and $[\,\underline{8},{}^3\!P_2]$ channels
\cite{cac,pko} and thus proportional to $M_r^{J/\psi}$.
In Section~\ref{s3}, we have seen that the fit result for $M_r^{J/\psi}$ is
greatly reduced by the inclusion of HO QCD effects on hadroproduction at low
$p_T$ \cite{can}.
This nourishes the hope that the apparent discrepancy between theory and
experiment at HERA may thus be alleviated at higher orders.

In Figs.~\ref{f4}(a)--(c), we take a first step towards a full NLO description
of the HERA data on inelastic $J/\psi$ photoproduction by repeating the
analyses of Figs.~\ref{f3}(a)--(c) using the HO-improved results of
Table~\ref{t1}, the NLO proton and photon PDF's specified in Section~\ref{s2},
and the two-loop formula for $\alpha_s$.
However, we do not include the NLO corrections to the partonic cross sections
in Eq.~(\ref{fus})
As already mentioned in Section~\ref{s2}, they are unknown, except for direct
photoproduction in the CSM \cite{kra}.
In the latter case, they were found to moderately increase the cross section,
by about 20\% \cite{kra}.
We observe that the HO-improved predictions tend to undershoot the data
leaving room for a substantial $K$ factor due to the missing NLO corrections
to the partonic cross sections in Eq.~(\ref{fus}).
Now, the colour-singlet contribution of direct photoproduction, which is well
under theoretical control \cite{kra}, is by far dominant, except in the
corners of phase space, at $z\alt0.15$ and $z\agt0.85$, where the colour-octet
contributions of resolved and direct photoproduction, respectively, take over.
It has been noticed \cite{kni,bak,rot} that special care must be exercised in 
the treatment of the latter contributions.
In the vicinity of the upper endpoint, HO nonperturbative contributions are
enhanced and lead to a breakdown of the NRQCD expansion, so that only an
average cross section over a sufficiently large region close to $z=1$ can be 
predicted \cite{rot}.
On the other hand, one my infer from the study of resolved $J/\psi$
photoproduction via fragmentation \cite{kni,bak} that the corresponding
partonic cross sections in Eq.~(\ref{fus}) are likely to receive significant
NLO corrections at low $z$.
Of course, we should also bear in mind that the predictions shown in
Figs.~\ref{f3} and \ref{f4} still suffer from considerable theoretical
uncertainties related to the choice of the scales $\mu$ and $M_f$, the PDF's,
and other input parameters such as $m_c$ and $\Lambda^{(4)}$ \cite{ben,kra}.
From these observations, we conclude that it is premature at this point to
speak about a discrepancy between the Tevatron \cite{abe,cdf} and HERA
\cite{aid,bre} data of inclusive $J/\psi$ production within the framework of
NRQCD \cite{bod}.

Finally, we present LO predictions for the inclusive photoproduction of
$\chi_{cJ}$ mesons with subsequent radiative feed down to $J/\psi$ mesons.
We work at LO in the fusion picture using the $\chi_{cJ}$ matrix elements of
Table~\ref{t2}.
In Figs.~\ref{f5}(a)--(c), we present our results for the $W$, $p_T^2$, and
$z$ distributions, respectively, under the same kinematical conditions as in
the $J/\psi$ case considered in Figs.~\ref{f3} and \ref{f4}.
Notice that direct photoproduction of $\chi_{cJ}$ mesons is prohibited at LO
in the CSM.
From Fig.~\ref{f5}(c), we observe that, except for $z\agt0.8$, the dominant
contribution arises from resolved photoproduction in the CSM.
Comparing Figs.~\ref{f5}(a) and (b) with Figs.~\ref{f3}(a) and (b), we see
that, after integration over $z$, this indirect source of $J/\psi$ mesons is
suppressed by roughly two orders of magnitude relative to prompt $J/\psi$
production.

\section{\label{s5}Conclusions}

We determined the $J/\psi$ colour-octet matrix elements which appear in the
NRQCD expansion \cite{bod} at leading order in $v$ by fitting the latest
Tevatron data of prompt $J/\psi$ hadroproduction \cite{cdf}.
We found that the result for the linear combination $M_r^{J/\psi}$ of
$\left\langle{\cal O}^{J/\psi}[\,\underline{8},{}^1\!S_0]\right\rangle$ and
$\left\langle{\cal O}^{J/\psi}[\,\underline{8},{}^3\!P_0]\right\rangle$ is
substantially reduced if the HO QCD effects due to the multiple emission of
gluons, which had been estimated by Monte Carlo techniques \cite{can}, are
taken into account.
As an important consequence, the intriguing excess of the LO NRQCD prediction
for inelastic $J/\psi$ photoproduction at $z$ close to unity \cite{cac,pko}
over the HERA measurements \cite{aid,bre} disappears.
We assess this finding as an indication that it is premature to proclaim an
experimental falsification of the NRQCD framework on the basis of the HERA 
data.
Although we believe that our analysis captures the main trend of the HO
improvement, we stress that it is still at an exploratory level, since a
number of ingredients which would be necessary for a fully consistent NLO
treatment of inclusive $J/\psi$ hadroproduction and photoproduction are still
missing.

\bigskip
\centerline{\bf ACKNOWLEDGMENTS}
\smallskip\noindent
One of us (G.K.) is grateful to the Theory Group of the
Werner-Heisenberg-Institut for the hospitality extended to him during a visit
when this paper was prepared.

\newpage

\vskip-6cm

\begin{figure}

\centerline{\bf FIGURE CAPTIONS}

\caption{\protect\label{f1}(a) LO and (b) HO-improved fits to the CDF data on
the inclusive hadroproduction of prompt $J/\psi$ mesons \protect\cite{cdf},
which come in the form of $d\sigma/dp_T$ integrated over $|\eta|<0.6$ as a
function of $p_T$.
The fragmentation results are only shown for $p_T\ge10.91$~GeV, while the 
fusion results are displayed over the full $p_T$ range.
\hskip9cm}

\vskip-.2cm

\caption{\protect\label{f2}LO fit in the fusion picture to the CDF data on the
inclusive hadroproduction of prompt $\chi_{cJ}$ mesons with radiative feed 
down to $J/\psi$ mesons \protect\cite{cdf}, which come in the form of
$d\sigma/dp_T$ integrated over $|\eta|<0.6$ as a function of $p_T$.
\hskip9cm}

\vskip-.2cm

\caption{\protect\label{f3}The ZEUS data on the inclusive photoproduction of
prompt $J/\psi$ mesons \protect\cite{bre}, which come in the form of
(a) $\sigma$ integrated over $0.4<z<0.8$ and $p_T^2>1$~GeV$^2$ as a function
of $W$,
(b) $d\sigma/dp_T^2$ integrated over $0.4<z<0.9$ and 50~GeV${}<W<180$~GeV as a
function of $p_T^2$, and
(c) $d\sigma/dz$ integrated over $p_T^2>1$~GeV$^2$ and 50~GeV${}<W<180$~GeV as
a function of $z$, are compared with the respective LO predictions.
The total results (solid lines) are built up by the
direct-photon colour-singlet (dotted lines),
direct-photon colour-octet (dashed lines),
resolved-photon colour-singlet (dot-dashed lines), and
resolved-photon colour-octet (dot-dot-dashed lines) contributions.
The respective H1 data \protect\cite{aid}, which refer to
30~GeV${}<W<150$~GeV, are also shown.
\hskip9cm}

\vskip-.2cm

\caption{\protect\label{f4}Same as in Fig.~\ref{f3}, but for the HO-improved
predictions.
\hskip9cm}

\vskip-.2cm

\caption{\protect\label{f5}LO predictions for the inclusive photoproduction of
prompt $\chi_{cJ}$ mesons with radiative feed down to $J/\psi$ mesons in the 
form of
(a) $\sigma$ integrated over $0.4<z<0.8$ and $p_T^2>1$~GeV$^2$ as a function
of $W$,
(b) $d\sigma/dp_T^2$ integrated over $0.4<z<0.9$ and 50~GeV${}<W<180$~GeV as a
function of $p_T^2$, and
(c) $d\sigma/dz$ integrated over $p_T^2>1$~GeV$^2$ and 50~GeV${}<W<180$~GeV as
a function of $z$.
The total results (solid lines) are built up by the
direct-photon colour-octet (dashed lines),
resolved-photon colour-singlet (dot-dashed lines), and
resolved-photon colour-octet (dot-dot-dashed lines) contributions.
Direct photoproduction is prohibited at LO in the CSM.
\hskip9cm}

\end{figure}

\newpage
\begin{figure}[ht]
\epsfig{figure=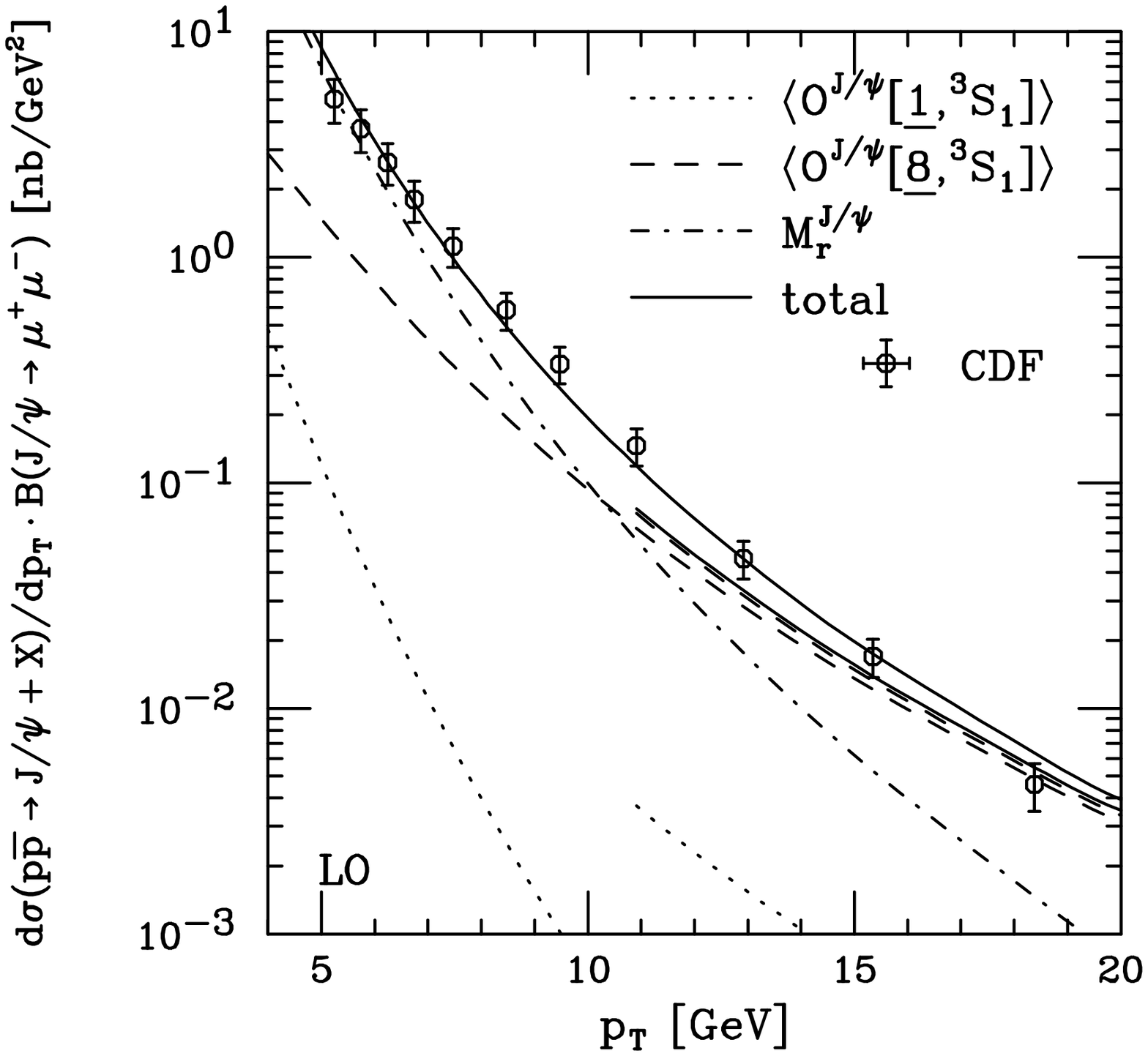,width=\textwidth}
\bigskip
\centerline{\Large\bf Fig.~1a}
\end{figure}

\newpage
\begin{figure}[ht]
\epsfig{figure=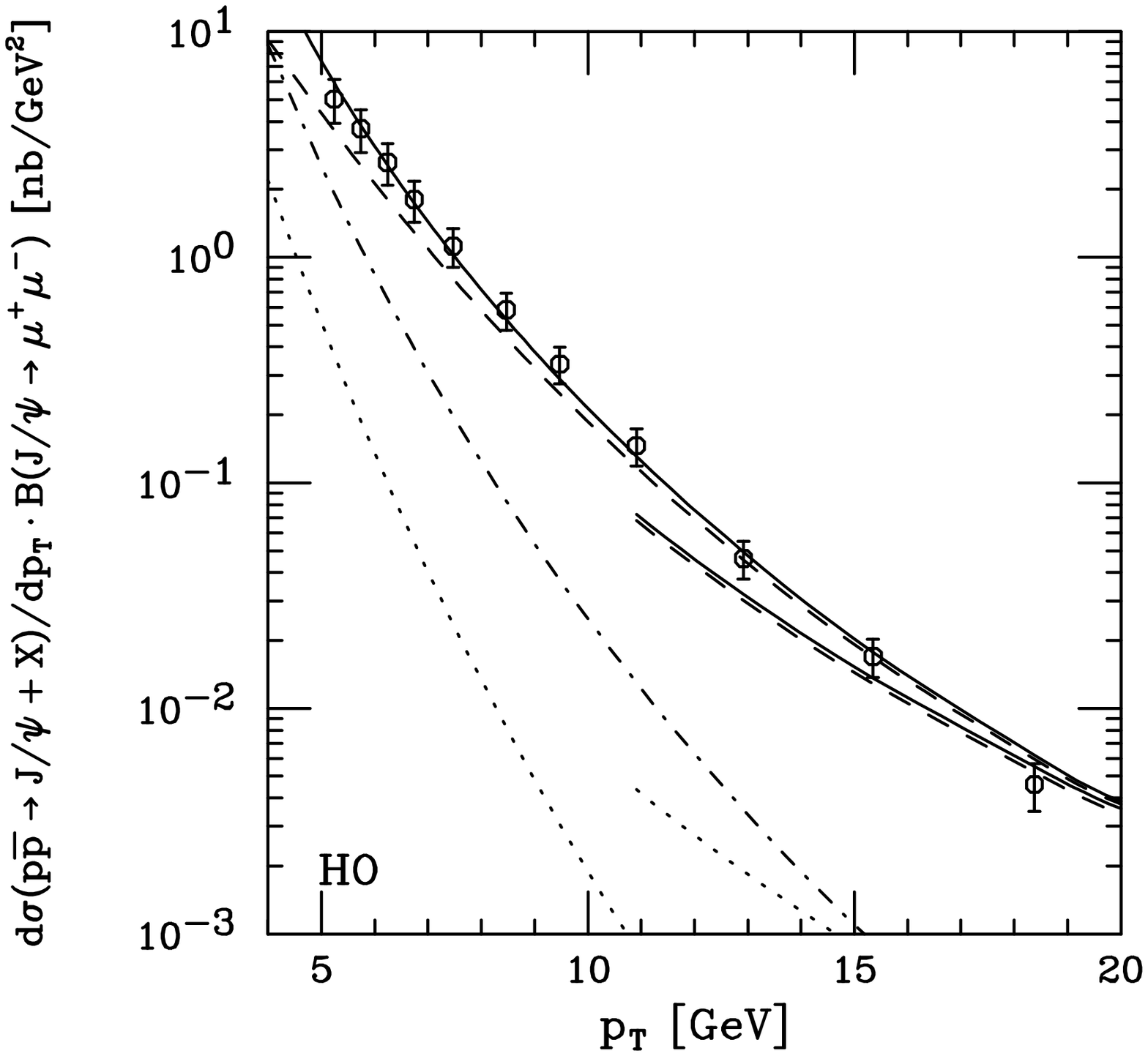,width=\textwidth}
\bigskip
\centerline{\Large\bf Fig.~1b}
\end{figure}

\newpage
\begin{figure}[ht]
\epsfig{figure=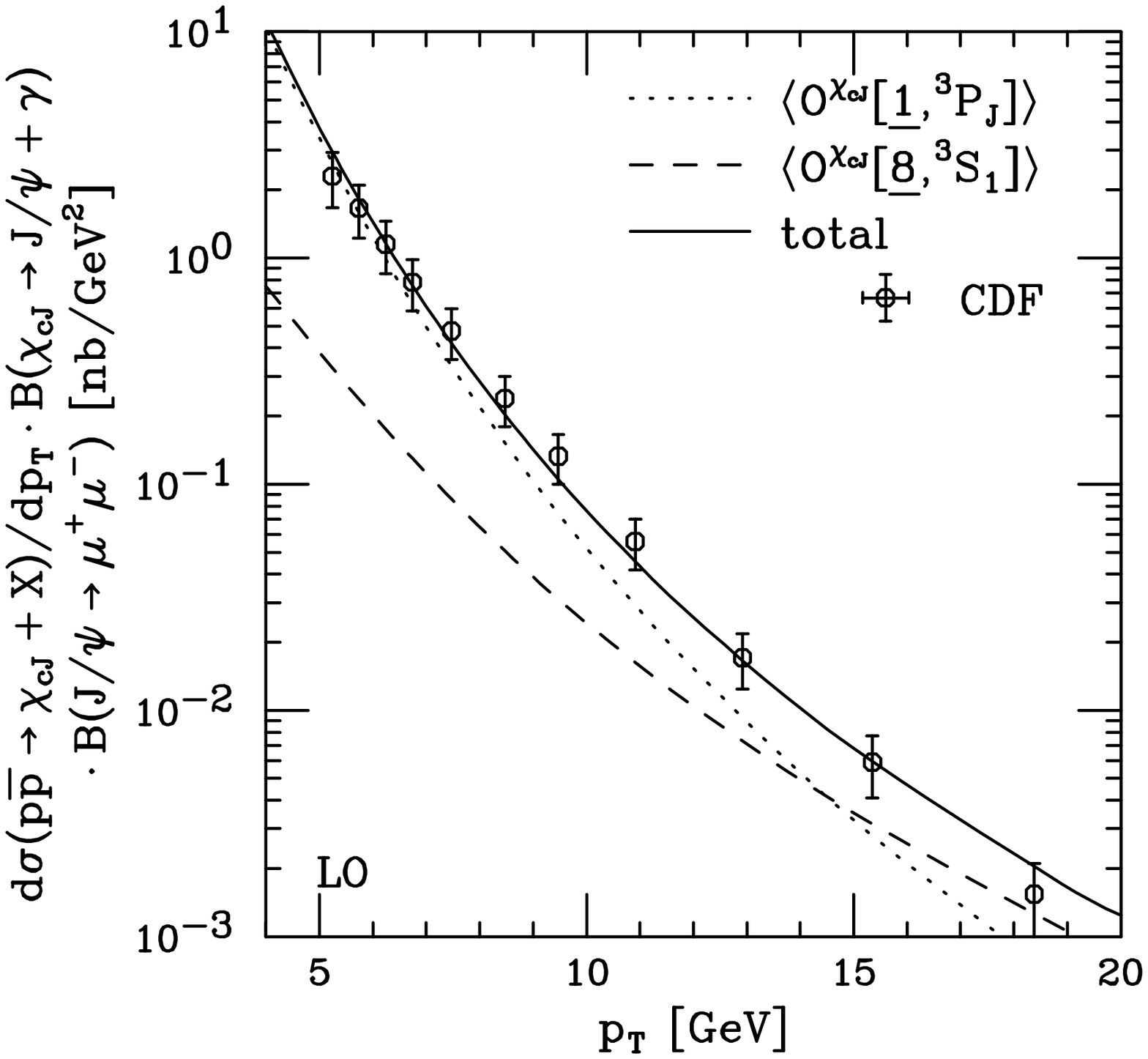,width=\textwidth}
\bigskip
\centerline{\Large\bf Fig.~2}
\end{figure}

\newpage
\begin{figure}[ht]
\epsfig{figure=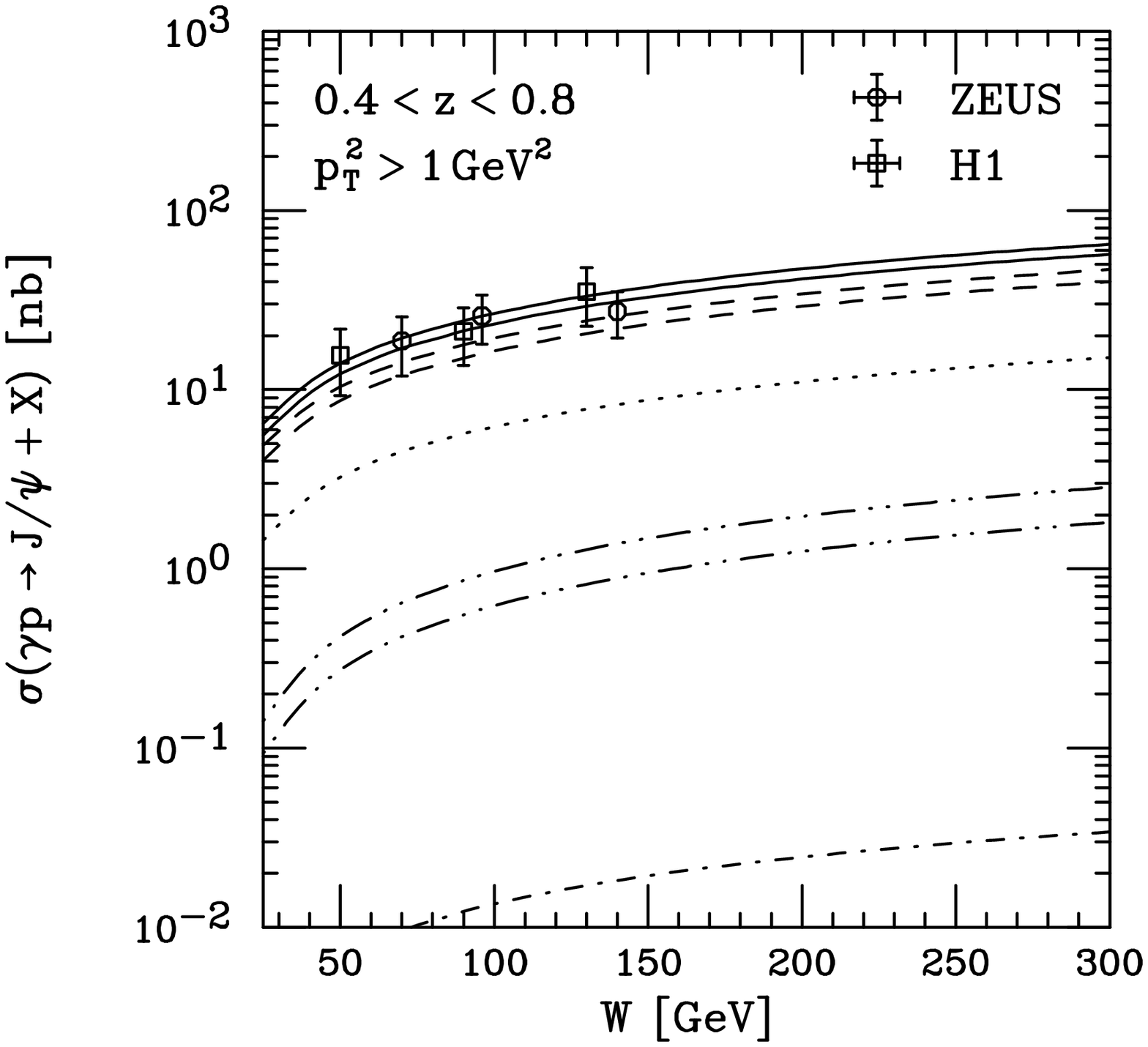,width=\textwidth}
\bigskip
\centerline{\Large\bf Fig.~3a}
\end{figure}

\newpage
\begin{figure}[ht]
\epsfig{figure=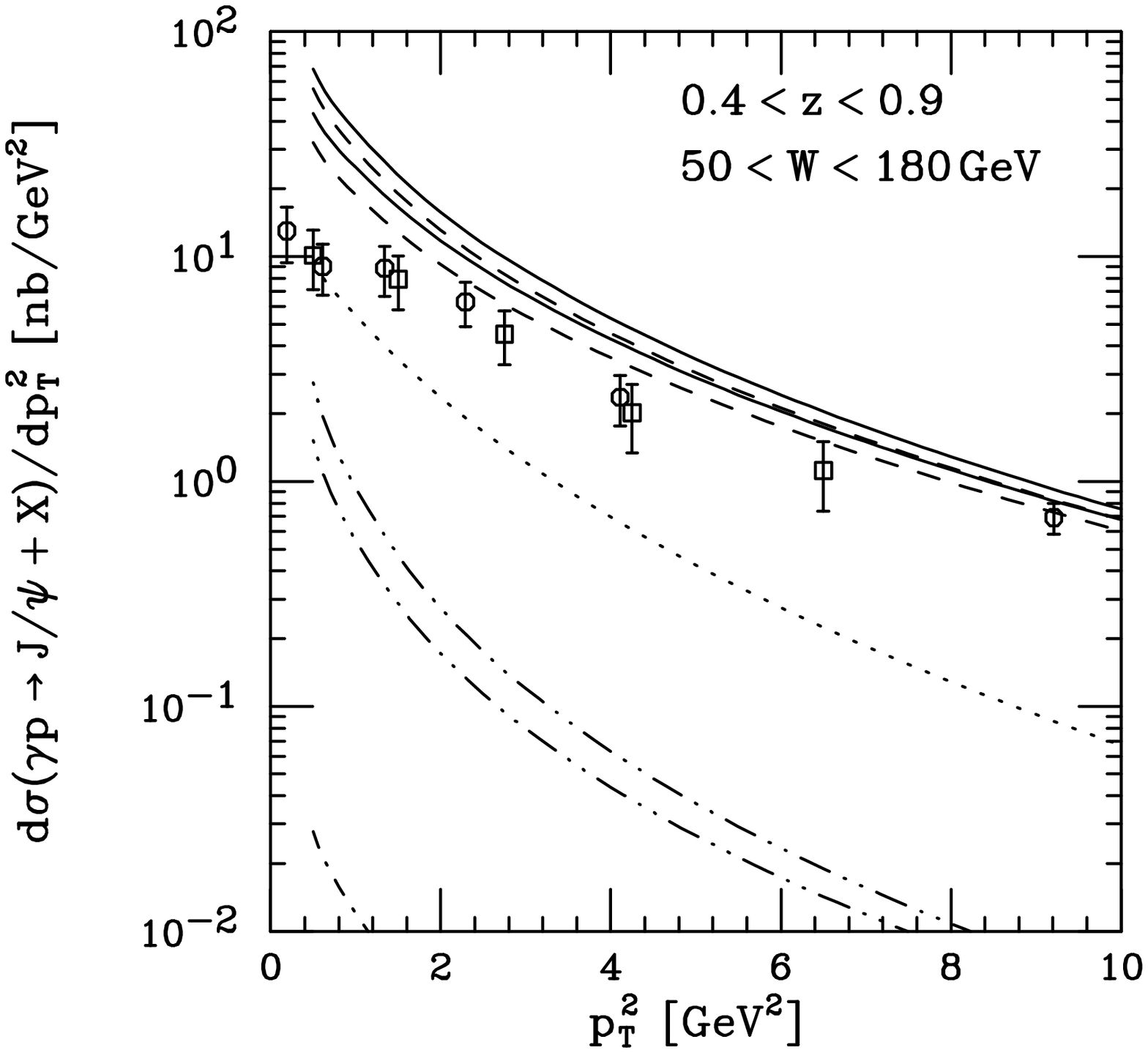,width=\textwidth}
\bigskip
\centerline{\Large\bf Fig.~3b}
\end{figure}

\newpage
\begin{figure}[ht]
\epsfig{figure=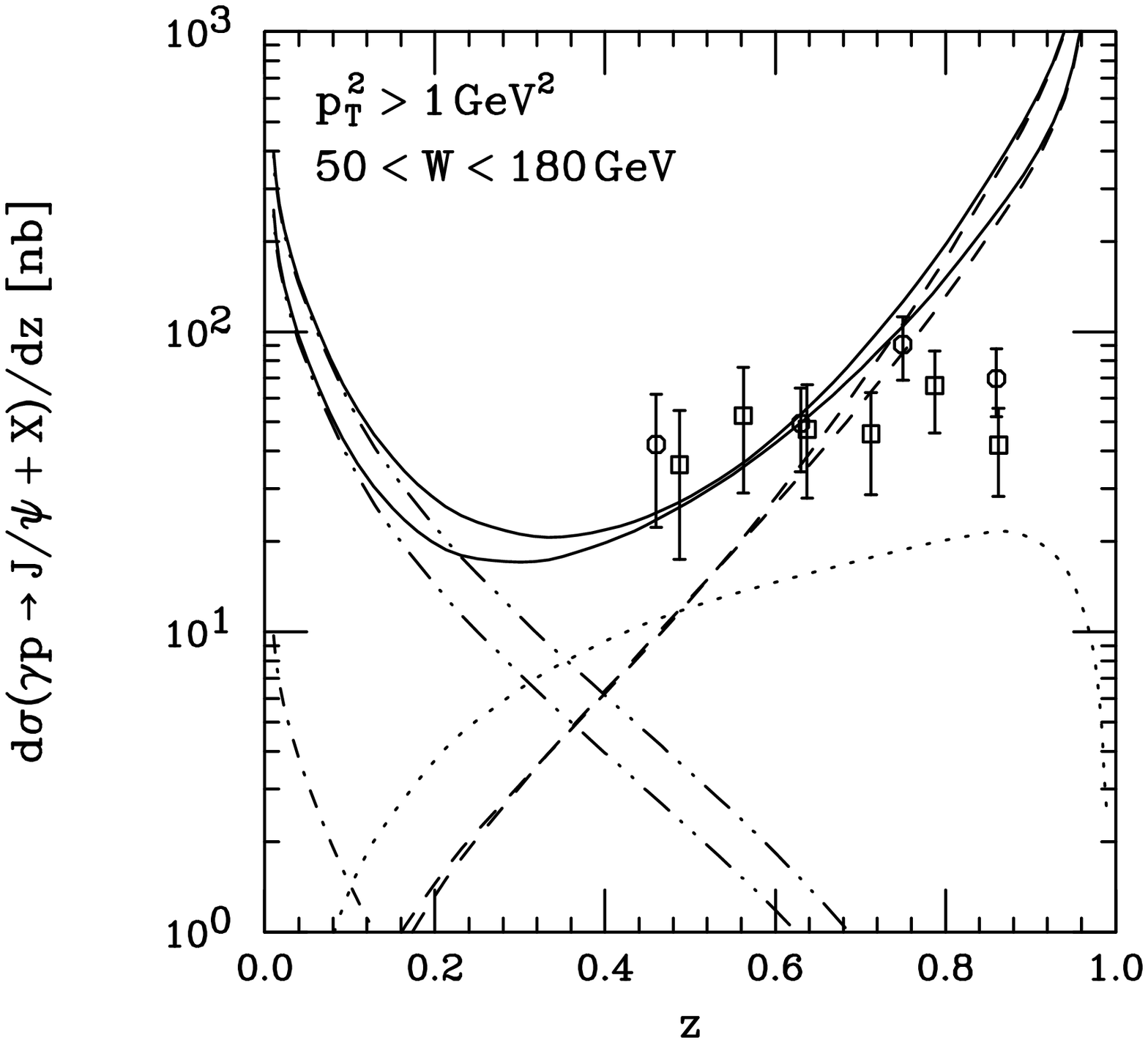,width=\textwidth}
\bigskip
\centerline{\Large\bf Fig.~3c}
\end{figure}

\newpage
\begin{figure}[ht]
\epsfig{figure=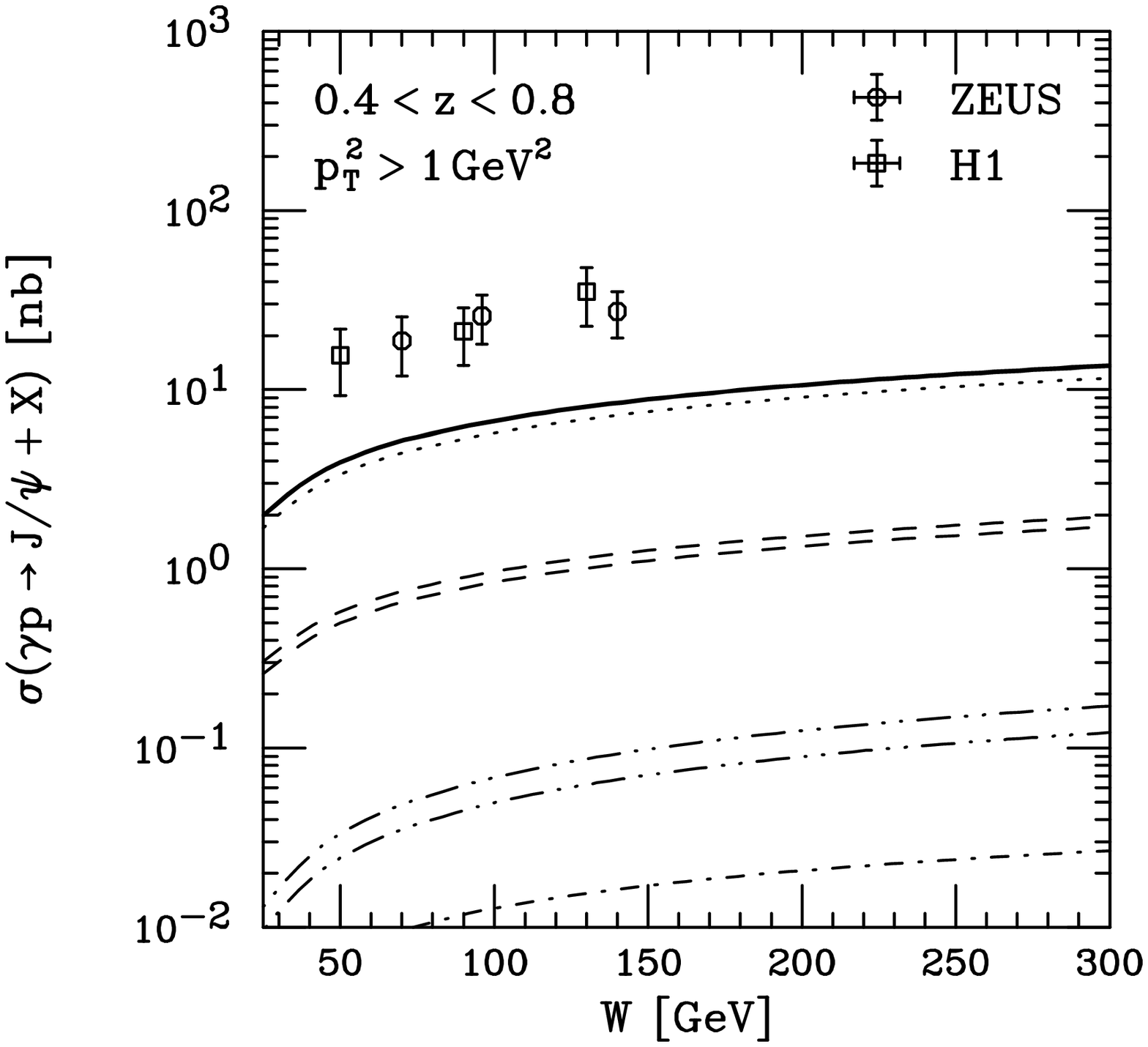,width=\textwidth}
\bigskip
\centerline{\Large\bf Fig.~4a}
\end{figure}

\newpage
\begin{figure}[ht]
\epsfig{figure=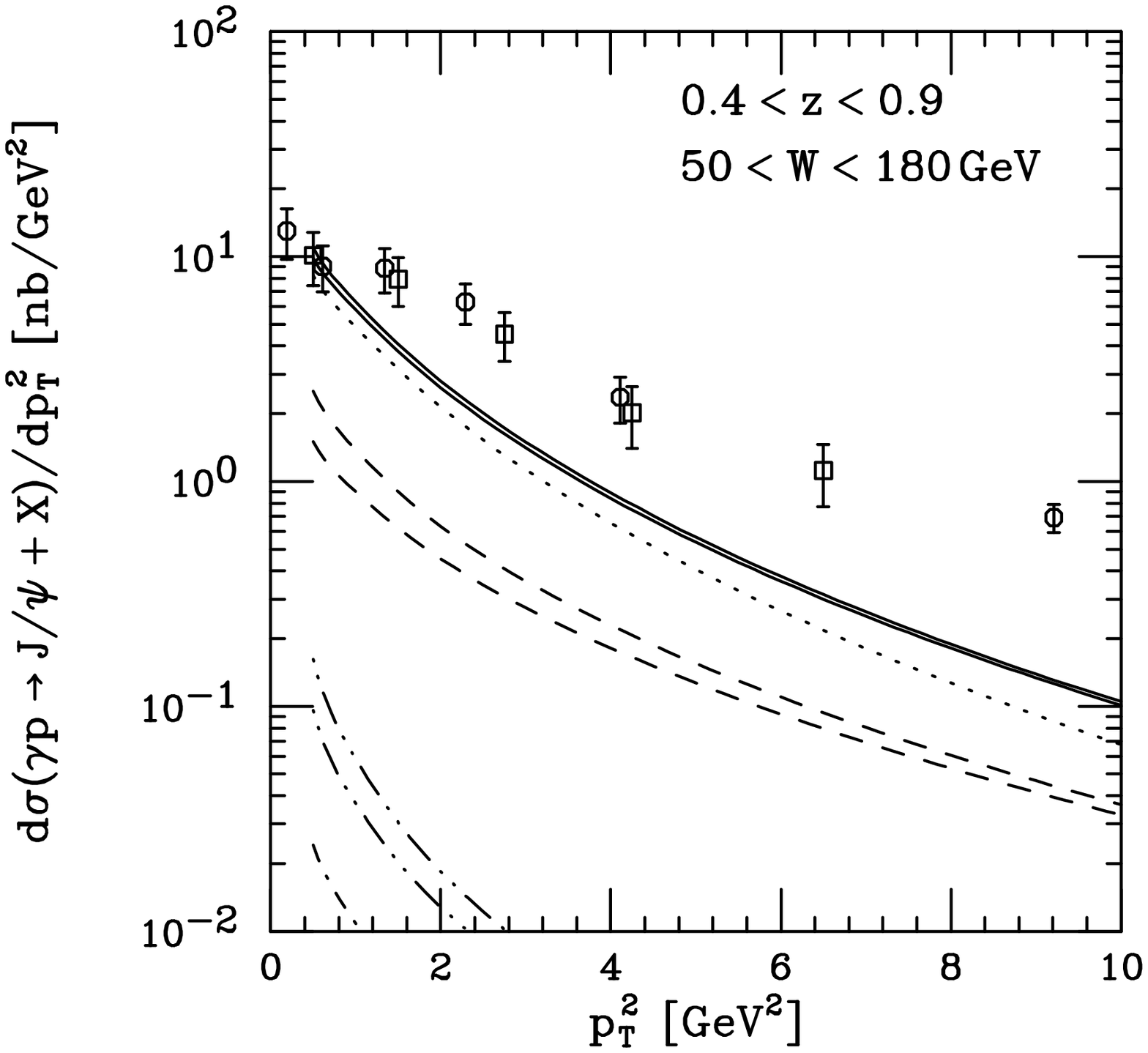,width=\textwidth}
\bigskip
\centerline{\Large\bf Fig.~4b}
\end{figure}

\newpage
\begin{figure}[ht]
\epsfig{figure=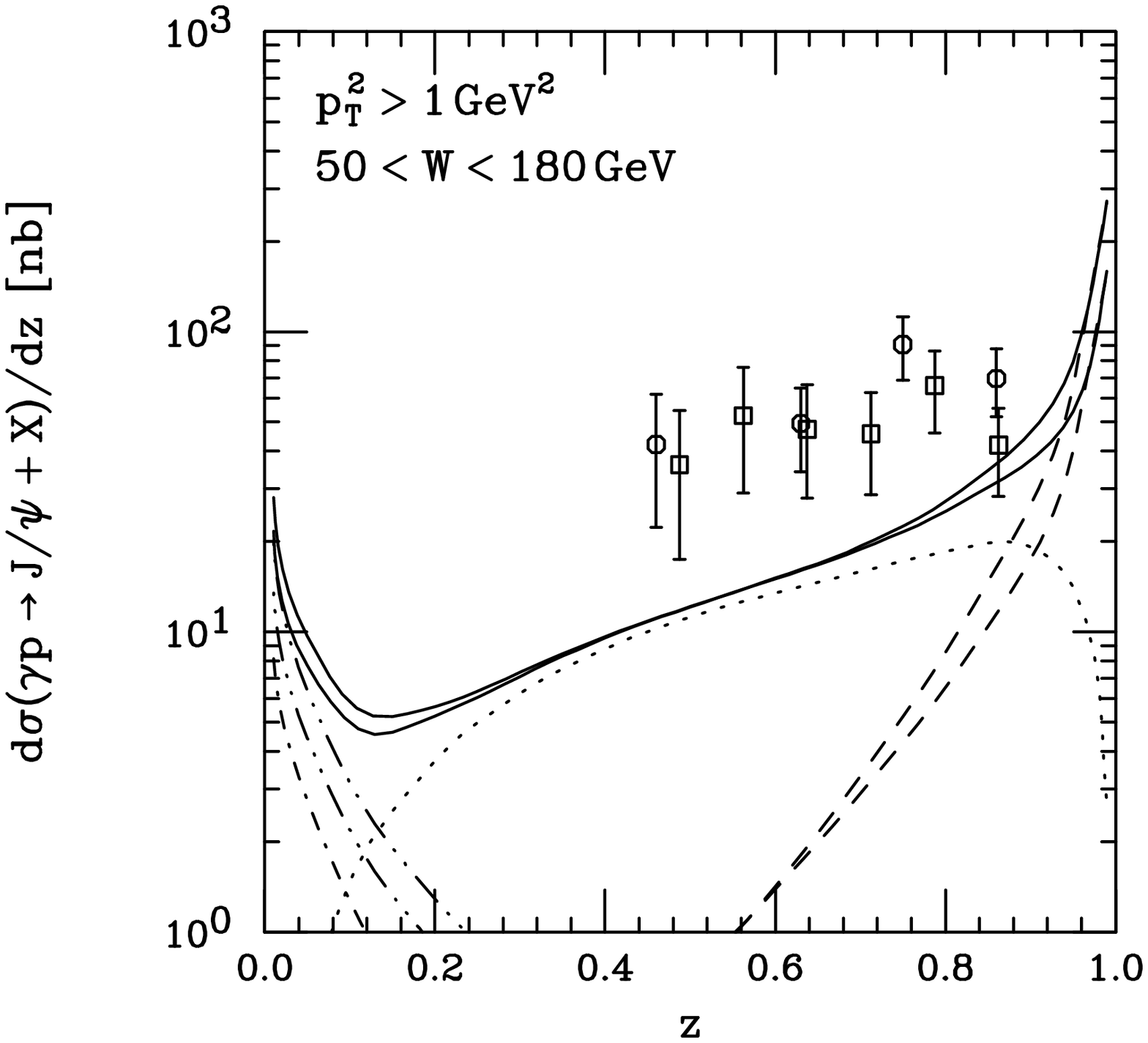,width=\textwidth}
\bigskip
\centerline{\Large\bf Fig.~4c}
\end{figure}

\newpage
\begin{figure}[ht]
\epsfig{figure=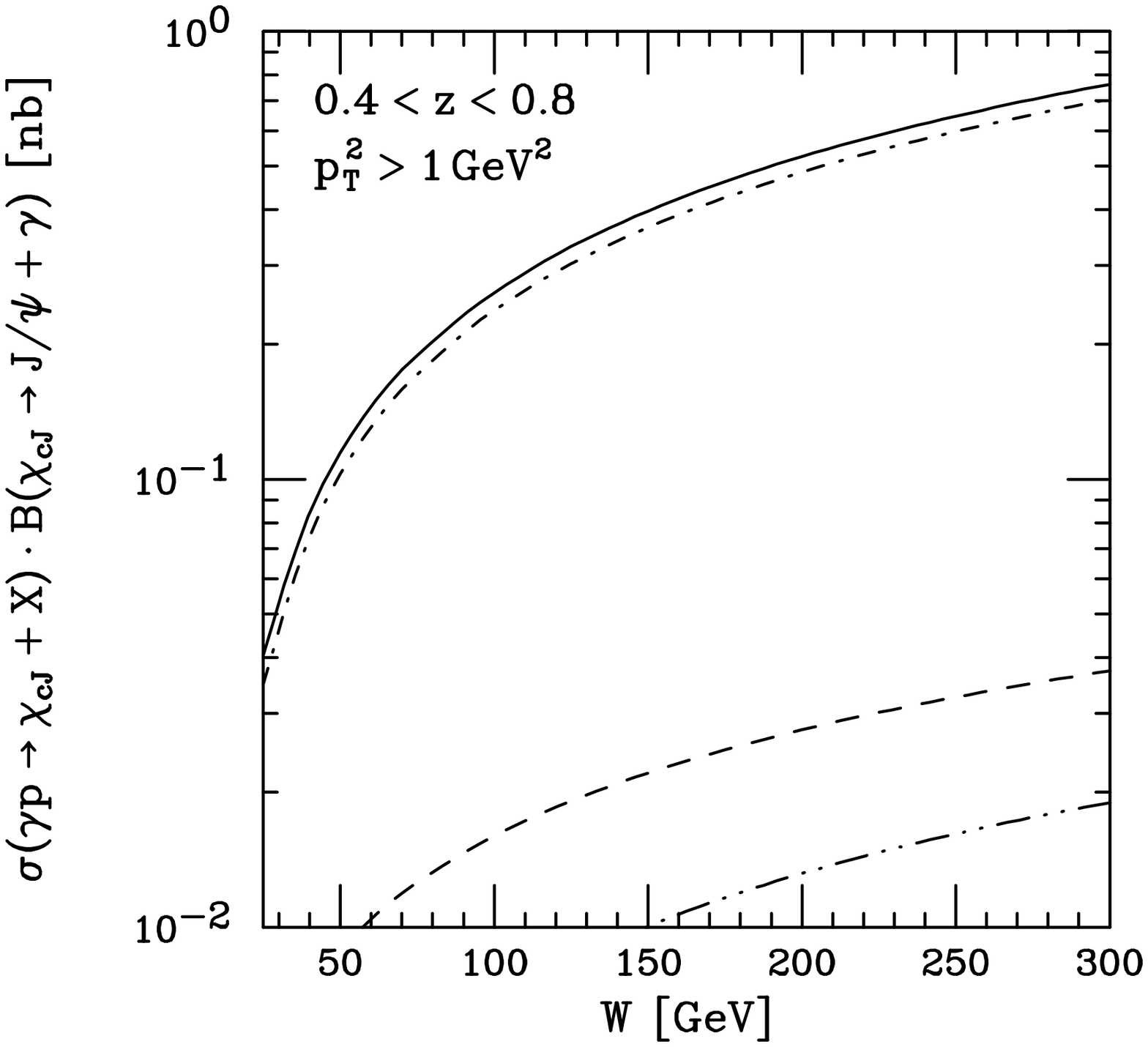,width=\textwidth}
\bigskip
\centerline{\Large\bf Fig.~5a}
\end{figure}

\newpage
\begin{figure}[ht]
\epsfig{figure=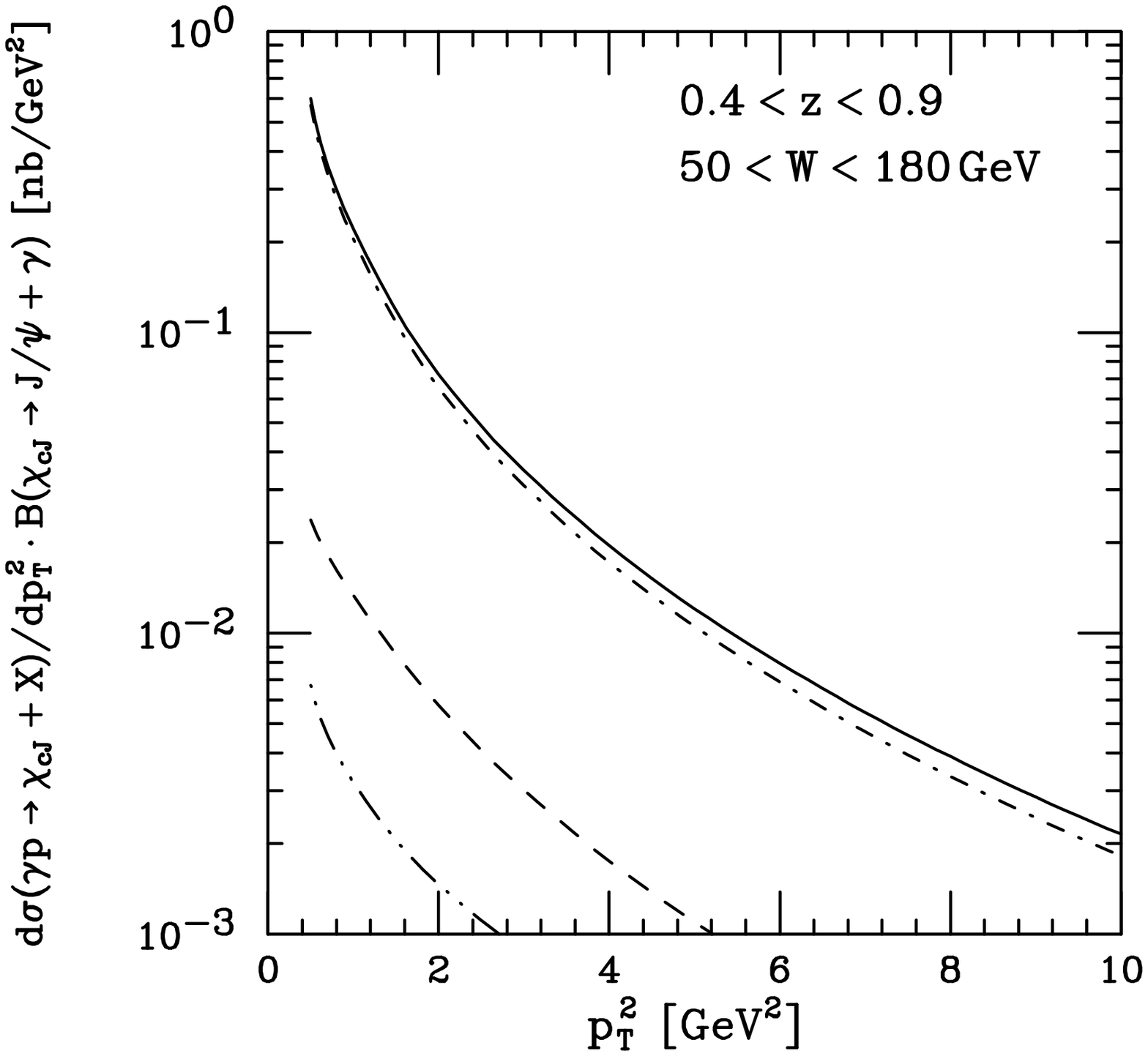,width=\textwidth}
\bigskip
\centerline{\Large\bf Fig.~5b}
\end{figure}

\newpage
\begin{figure}[ht]
\epsfig{figure=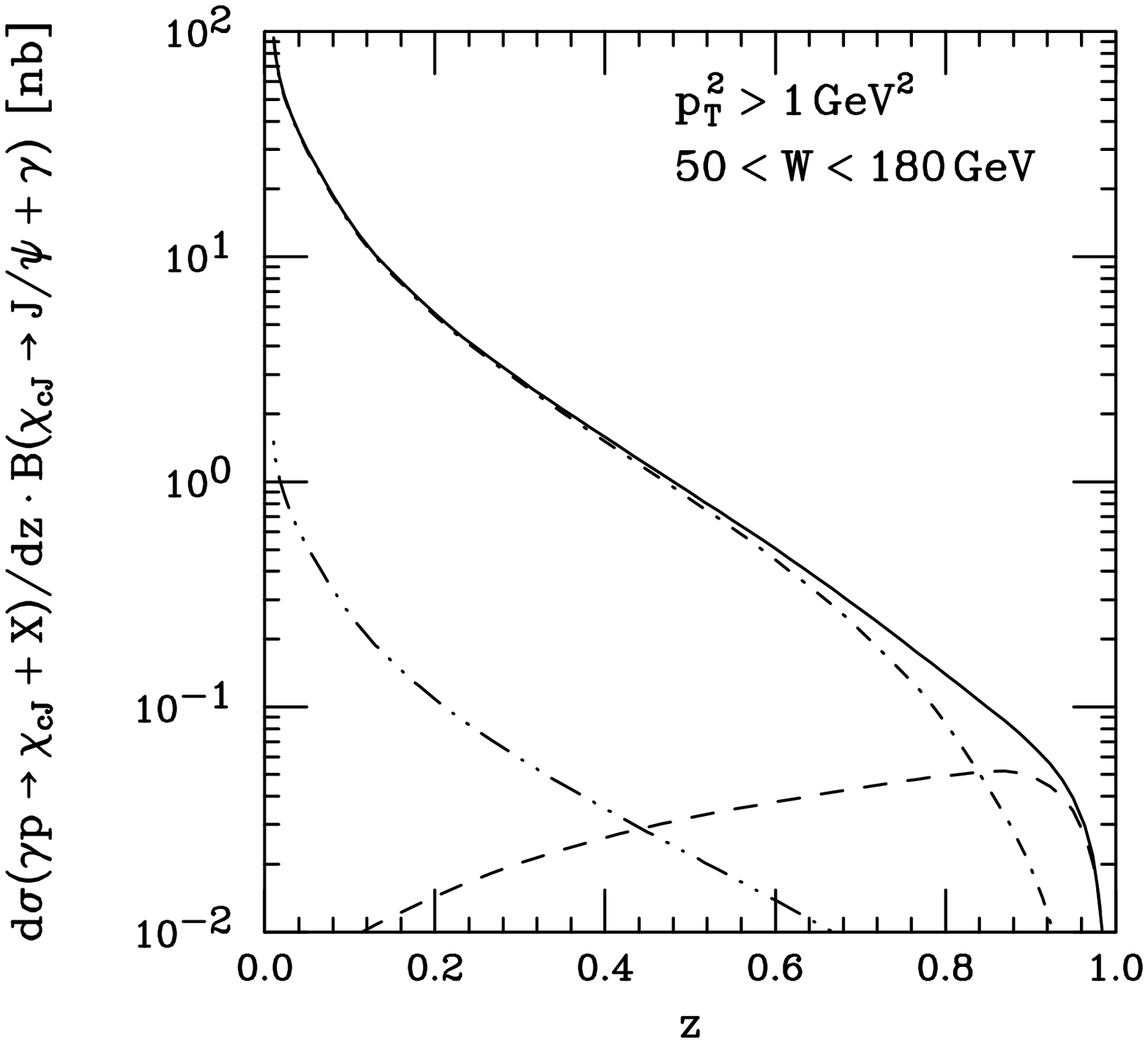,width=\textwidth}
\bigskip
\centerline{\Large\bf Fig.~5c}
\end{figure}

\end{document}